\RequirePackage{amsmath}
\documentclass [twocolumn]{svjour3}
\usepackage[english]{babel}
\usepackage{booktabs}
\usepackage[numbers,sort]{natbib}
\usepackage{bm}
\usepackage{graphicx}
\usepackage[utf8]{inputenc}
\usepackage[T1]{fontenc}
\usepackage[switch]{lineno} 
\usepackage[hidelinks]{hyperref}
\usepackage[inline]{enumitem}

\begin{document}

\title{Simulation and evaluation of cloud storage caching for data intensive science}

\author{Tobias Wegner \and Mario Lassnig \and Peer Ueberholz \and Christian Zeitnitz}

\institute{Tobias Wegner \and Mario Lassnig \at European Organization for Nuclear Research (CERN)\\
              \email{tobias.wegner@cern.ch}
           \and Peer Ueberholz \at
              Hochschule Niederrhein\\
              \email{peer.ueberholz@hs-niederrhein.de}
           \and Christian Zeitnitz \at
              University of Wuppertal\\
              \email{zeitnitz@uni-wuppertal.de}
}

\date{Received: 1970-01-01 / Accepted: 1970-01-01}

\maketitle
\begin{abstract}
A common task in scientific computing is the derivation of data.
This workflow extracts the most important information from large
input data and stores it in smaller derived data objects. The
derived data objects can then be used for further analysis tasks.\\
Typically, those workflows use distributed storage and computing
resources. A straightforward configuration of storage media would
be low cost tape storage and higher cost disk storage. The large,
infrequently accessed input data is stored on tape storage. The
smaller, frequently accessed derived data is stored on disk storage.
In a best case scenario, the large input data is only accessed very
infrequently and in a well planned pattern. However, practice shows
that often the data has to be processed continuously and
unpredictably. This can significantly reduce tape storage performance.
A common approach to counter this is storing copies of the large
input data on disk storage.\\
This contribution evaluates an approach that uses cloud storage
resources to serve as a flexible cache or buffer depending on the
computational workflow. The proposed model is elaborated for the
case of continuously processed data. For the evaluation, a
simulation was developed, which can be used to evaluate models
related to storage and network resources.\\
We show that using commercial cloud storage can reduce the
on-premises disk storage requirements, while maintaining an
equal throughput of jobs. Moreover, the key metrics of the
model are discussed and an approach is described that uses
the simulation to assist with the decision process of using
commercial cloud storage. The goal is to investigate
approaches and propose new evaluation methods to overcome the
future data challenges.
\end{abstract}

\setcounter{tocdepth}{2}


\section{Introduction}
\label{sec:introduction}
Modern scientific experiments, such as ATLAS \cite{atlas:paper}, CMS \cite{cms},
Vera Rubin Observatory \cite{LSST}, or SKA \cite{ska}, generate very large data
samples. The large volume of these data samples typically requires distributed
computing and storage resources \cite{GRIDHEP04, WFMSGRID05, WLCG06} to process
and store the data \cite{ADC17, PROD17, DDM17}. These resources are pooled in
data centres and consist of different kinds of storage media with different
Quality-of-Service (QoS) characteristics. Straightforward QoS deployments
typically consist of disk and tape storage \cite{TAPEPLACEMENT06} targeting
certain experiment needs like access latency, access pattern, throughput, or
cost.\\
Tape storage typically comes with a high access latency but low cost per volume
ratio compared to disk storage \cite{STORAGECOST07, YU_2011}. A common use case
of tape storage is the archiving and preservation of infrequently required data.
In addition, the throughput of tape storage strongly depends on the data placement
and a predictable access pattern \cite{TAPEPLACEMENT06, YU_2011}. This makes tape
storage a preferable storage type from a cost-oriented perspective, while disk
storage is a preferred storage from a performance-oriented perspective.

In general, data intensive computing workflows require the use of
performance-oriented storage, such as disks, because of the lower response time
and higher throughput in concurrent and random access mode. This often results
in maintaining at least one persistent copy of each input file on a disk system.\\
A typical workflow in scientific computing is the derivation of data to extract
only the most important information and reduce the data volume, i.e., the volume
of the input data of the derivation workflow is usually much larger than the volume
of the output data. These workflows could be scheduled to be infrequently executed
in bulk processing campaigns, e.g., whenever a new derivation software version is
released. In this way, input files have to be read only once per campaign. Given this
case, the input data would be preferably stored solely on tape. However, for larger
collaborations, such as ATLAS, it is challenging to optimally organise those workflows
into campaigns. For this reason, there are continuous derivation workflows that read
input data more frequently.\\
To allow a continuous derivation workflows to have a proper throughput of input data,
a common solution is to keep at least one copy of the vast majority of input files
on a disk storage system. For example, almost all input data for the production of
derivation data for ATLAS have one persistent copy on both tape and disk storage.

One approach that tries to take advantage of the infrequent usage of the input data
for derivation production campaigns is implemented in the \emph{data carousel}
model \cite{DATACAROUSEL20}. The concept of the data carousel model is to transfer
the input data from tape storage to disk storage, start processing the data, and
continuously replace the data that has been processed by new data coming from tape.\\
Using this model only a limited number of input files are required on disk storage
at any one time. This allows removing the permanent copies of the input files from
the disk storage system and storing the input files solely on tape storage. In this
way, disk storage requirements are reduced to save cost or provide disk storage for
other types of data.\\
The data carousel model was developed to improve storage usage and tape throughput
if the derivation workload is structured into campaigns so that the input files are
required only once per campaign. For the continuous derivation workflow, the input
files are accessed frequently which would result in using the tape storage in
concurrent random access mode and thus, significantly reduce the tape throughput
and performance.

To reduce these limitations the data carousel model can be combined with the
\emph{Hot/Cold Storage} model. The Hot/Cold Storage model categorises the storage
in three different QoS categories: a large archival storage, a medium sized cold
storage, and small hot storage. The data is migrated between hot and cold storage
based on a popularity metric. The concept is to use the cold storage as buffer for
the archival storage to improve its throughput or as cache for the hot storage to
reduce the number of re-transfers from the archival storage.\\
The VR Observatory decided to use cloud resources from Google as interim
data facility in 2020/23 \cite{vrogcp}. ATLAS is evaluating different approaches of
adopting commercial cloud resources \cite{dataocean}. The Hot/Cold Storage model has
been developed assuming that commercial cloud storage for at least one of the three
storage categories will be used.\\
In this contribution, we describe the \emph{Hot/Cold Data Carousel} (HCDC) model,
which is a combination of the Hot/Cold Storage model and the data carousel model.
Furthermore, we present a simulation to evaluate the HCDC model. The HCDC model aims
to minimise the disk storage required in derivation campaigns as achieved by the data
carousel, while mitigating the negative impact on the tape storage throughput for
continuous derivation workflows.

Section \ref{sec:fundamentals} starts by defining the basic terminology.
Subsequently the main assumptions that were made for the evaluation are
described. Finally the Hot/Cold Storage and data carousel model are explained
in more detail. Section \ref{sec:hcdc} describes the HCDC model and lists
possible variations of it. Section \ref{sec:simulation} starts by describing the
architecture of the simulation software that was developed for the evaluation
of the HCDC model. To validate the simulation a simplified scenario was
simulated and evaluated which is described following the architecture
description. Finally it is explained how the HCDC model was implemented in
the simulation and which parameters were used.
Section \ref{sec:evaluation} shows the results of the simulation of the HCDC
model and how the results were evaluated.
We conclude in Section \ref{sec:conclusion} with a summary and an outlook on
future work.

\section{Fundamentals}
\label{sec:fundamentals}
Both the Hot/Cold Storage model and the data carousel model were developed based on
the computing infrastructure of the ATLAS experiment which obtains its resources
from the \emph{Worldwide LHC Computing Grid} (WLCG) \cite{WLCG06, lhcopn}.
As mentioned before, globally distributed storage and compute resources
are pooled in data centres. In the context of grid resources, those data centres
are called \emph{sites}. The storage resources of a site are logically grouped in
\emph{storage elements}. Storage elements could differ in the attributes of their
underlying physical storage media or simply by the type of data they store.

The Hot/Cold Storage model considers the usage of commercial cloud resources.
Specifically, the development and evaluation of the model have been performed
considering resources from the \emph{Google Cloud Platform} (GCP) \cite{gcp}.
\emph{Google Cloud Storage} (GCS) denotes only the storage resources of the GCP.\\
Analogously to grid resources, clouds usually provide their resources pooled in
\emph{regions}, which represent the data centres. Storage resources are logically
divided into \emph{buckets}. In contrast to storage elements, cloud providers often
allow buckets being multi-regional. This means the data stored in a
multi-regional bucket is transparently replicated across multiple data centres.
In the scope of the earlier mentioned Data Ocean project \cite{dataocean},
the possibility of a scalable, globally accessible bucket was discussed
with Google. Originally, the HCDC model was developed based on such a bucket.
However, for the presented implementation the bucket is not required to be
globally accessible. The possible options to implement such a bucket still
have to be investigated. A straightforward approach would be a transparent
replication of the data in the bucket to regional data centres.

The HCDC model was evaluated based on the derivation workflow described in the
introduction. The proposed model requires the workflows to be executed in
one of two modes. The first mode assumes the derivation workflow is organised
in predictable campaigns resulting in an infrequent requirement of the input
data. The other mode assumes the derivation to run continuously which leads to a
more frequent and less predictable demand for the input data. The existence of
a popularity metric, such as the access frequency of a file, is assumed for the
continuous mode.

\subsection{Data Carousel model}
\label{sec:fundamentals_dc}
The data carousel model aims at reducing the usage of performance-oriented
storage like disk storage and prefers the usage of low-cost storage like tape
storage. This is particularly applicable to workflows whose input data is
required rather infrequently or for data with an easily-predicted access
pattern. As is typically the case in scientific computing, the data carousel
model requires that the workload can be divided into discrete units.\\
A derivation campaign starts with the definition of the workload. In ATLAS
this is done by a production team creating \emph{tasks} in the production
system \cite{prodsys}. The production system coordinates with the data
management system \cite{DDM17} and the workflow management system \cite{panda}
the transfer of the data from tape storage to disk storage and the start of
the data processing.\\
In data carousel mode the input data is solely stored on tape. When the
derivation campaign is defined and the data to process is determined, a
\emph{sliding window} is created. The sliding window has a specific
size, e.g., the size of a given percentage of the total input data. The
data that is required for processing must allocate space in the sliding
window. After a successful allocation, the data can be transferred from
tape to disk storage. When enough data has been transferred, the workload
can start to process the data. The data is downloaded from the disk
storage to the worker nodes where it is processed by the derivation
software. When the processing of the data completes, the corresponding
data is deleted from the disk storage and deallocated in the sliding
window. Using this approach, only disk storage equal to the sliding
window size is required at any one time.\\
The possible size of the sliding window is limited by the following
parameters:
\begin{itemize}
    \item available storage for the sliding window
    \item volume of input data to process
    \item throughput and latency of the tape storage
    \item time between start of transfers and start of workloads
    \item available computing resources
\end{itemize}
The minimal and maximal size of the sliding window depends on the
available storage and the volume of the required input data. Typically,
the volume of the input data is larger than the storage available for
the sliding window. Thus, the temporary storage available for processing
is the limit rather than the volume of the input data. The window size
must be large enough to hold all the input data for all currently
running jobs.\\
Another potential limitation of the size of the sliding window is
given by the throughput from the tape storage to the disk storage
and the time it takes to process the data. For example, if the
the throughput from the tape to the disk storage is the bottleneck,
a very small sliding window size would be sufficient. The reason is
that a large window could not be filled up.

\subsection{Hot/Cold Storage model}
\label{sec:fundamentals_hcs}
\begin{figure}[t]
  \centering
  \includegraphics*[width=\columnwidth]{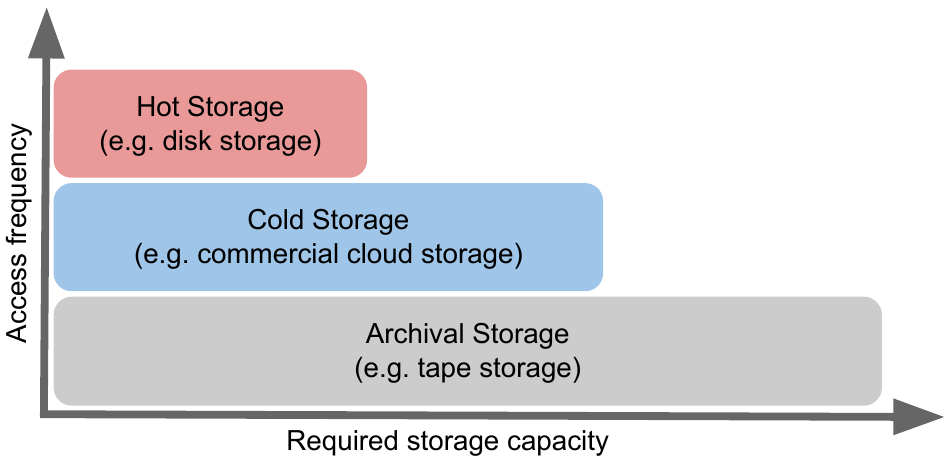}
  \caption{Hot/Cold storage model. One replica of each file is stored on archival
  storage. Files are migrated between cold and hot storage based on a popularity metric.}
  \label{fig:hotcold_storage_model}
\end{figure}

As shown in Figure \ref{fig:hotcold_storage_model}, the Hot/Cold Storage model
divides the storage into hot storage, cold storage, and archival storage. The
main dimensions in which the requirements to the storage categories differ are
the storage capacity and a popularity metric, such as the access frequency of
the data estimated by the number of times a file has been used.\\
Hot storage requires a small capacity to store only the most frequently accessed
data. Optimally, the hot storage would be located close to the computing resources
that require the data. Regarding the QoS properties, the storage implementing the
hot storage must provide good performance in terms of throughput and access
latency, especially in concurrent and random access mode.\\
Cold storage requires a larger storage capacity than hot storage. There
are two use cases for cold storage. First, it can be used as a temporary
buffer when the hot storage is full. In this case, cold storage accepts
data from archival storage that is required or is likely to be required
based on the popularity metric. Second, it can be used as a cache between
the archival and the hot storage. In this case, the cold storage caches
the data from the hot storage that is no longer required on the hot
storage but is likely to be required again in the short term.\\
Archival storage requires the largest capacity. The QoS properties of
archival storage typically describe a higher access latency and
significantly lower throughput for concurrent and random access mode.
The Hot/Cold Storage model assumes that at least one replica of each
file is kept in an archival storage.\\
Various approaches are possible to use together the different storage
categories. The data on hot storage is replaced very frequently. The
data is preferably transferred from a cold to a hot storage. If the
required data is not available on a cold storage it is transferred
from an archival storage.\\
In the implemented variation of the Hot/Cold Storage model, the
required data that is not available on the cold storage is directly
transferred from the archival to the hot storage. Prior to the
deletion of data at the hot storage, the data is replicated to the
cold storage. Alternatively, the required data could firstly be
transferred from the archival to the cold storage and then be
transferred to the hot storage. This would result in less delay
for the deletion because the data does not have to be transferred
to the cold storage first. However, it would increase the waiting
time for the required data.\\
Another point is how the allocation and deallocation of cold storage
are managed. Ensuring the existence of hot storage data on cold
storage prior to their deletion, requires either an unlimited cold
storage capacity or a deletion strategy of cold storage data. Another
approach would be to set a threshold based on the popularity metric
and only allow transferring data to the cold storage that have a
certain popularity. This threshold could be used to improve the
hit/miss ratio when using the cold storage as cache.

\section{HCDC model}
\label{sec:hcdc}
The HCDC model joins the Hot/Cold Storage model with the data carousel model. The
model includes two types of resource providers. First, the institution or company
using the model. Resources of this provider are persistently available and can be
used independently of the cost, e.g., the institutions that are associated with the
ATLAS experiment provide a pledged amount of resources.\\
Second, commercial cloud providers providing resources with typical cloud behaviour,
i.e., they can be allocated and deallocated on demand to an arbitrary extent.
However, they induce different kinds of costs. Regarding storage resources, there
are at least costs for the stored volume per time, depending on the storage type.
Furthermore, costs for operations like writing, reading, deleting, or changing
metadata of files, are typically charged. Another cost factor is the network
traffic. Usually, cloud providers only charge egress and traffic between different
regions within the cloud. The amount charged for traffic within the same cloud
depends on the source and destination endpoints. Egress traffic out of the cloud
to the internet is the most expensive.

The Hot/Cold Storage model and the data carousel model can be combined in different
variations. The first consideration is whether the derivation workload operates in
continuous mode or is executed in campaigns. For organised campaigns, the data
carousel part is expected to be most effective, whereas the Hot/Cold Storage part
should be less impactful. As described in Section \ref{sec:fundamentals}, this is
because for organised campaigns the data is required infrequently and the order to
read the files can be well planned before starting the workload. In this case, the
potential benefit from the  Hot/Cold Storage model part would be to use the cold
storage as a prefetching area. This could avoid the throughput from the archival
storage from becoming a bottleneck after the start of the campaign. In continuous
production mode, the data is required more frequently and less predictably. This
leads to the expectation that the data carousel part is less impactful. However,
the Hot/Cold Storage part should become more important in providing a cache for
the processed data and reduce the data access on archival storage.\\
Another point to consider is which storage category of the Hot/Cold Storage model is
represented by the cloud storage. This depends on the different storage types
offered by the cloud provider and should be decided based on the QoS properties 
described in Section \ref{sec:fundamentals_hcs}.\\
Using the cloud storage as an archive, would necessitate the acquisition of the largest
amount of cloud storage among all three storage categories. However, some cloud
providers offer different storage types with different pricing policies. For
example, GCS offers storage types with a low cost per volume ratio, but a higher
cost per access ratio. Since the archival storage category is used for the least
frequently accessed data, the higher cost per access ratio could be negligible and
the lower cost per volume ratio could be beneficial.\\
Using the cloud storage as hot storage, would reduce the required cloud storage
volume. Since hot storage requires high performance storage, the cost per volume
for the storage type would typically be higher. Furthermore, the egress cost would
be very high if the data is processed outside of the cloud because the hot storage
contains the most popular data.\\
%
Cold storage has a more flexible volume requirement depending on the popularity
metric and the available hot storage. With cloud storage as cold storage, the
egress cost would depend on the number of reusages of the data and the available
amount of hot storage.\\
Using the cloud for cold storage enables the sliding window of the data carousel
model to become dynamic. That means that, in case hot storage is full, the sliding
window can be extended by using cold storage to keep an optimal throughput from the
archival storage.

\begin{figure}[t]
  \centering
  \includegraphics*[width=0.8\columnwidth]{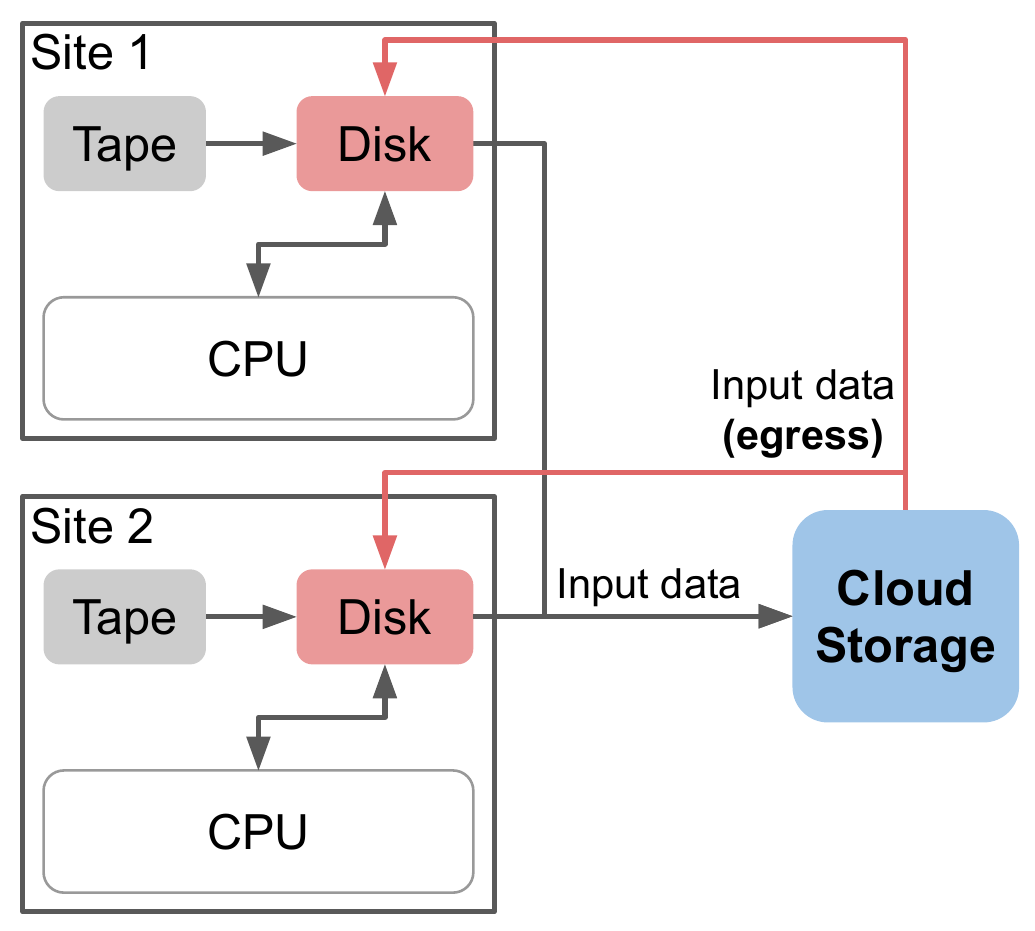}
  \caption{Schematic illustration of the combination of the Hot/Cold Storage
  model and the data carousel model with two sites.}
  \label{fig:model_hcdc}
\end{figure}

Figure \ref{fig:model_hcdc} illustrates how the Hot/Cold Storage model could be
combined with the data carousel model, for the simple case of two sites. The
storage categories of the Hot/Cold Storage model part are assigned
straightforwardly using site tape storage as archival storage, cloud as cold
storage, and site disk storage as hot storage.\\
To read data from a tape system of the grid, the data always has to be
transferred to a small disk storage buffer area first. The graphic shows the
disk storage summarised in a single box per site. In this model, the data on
tape is read only and there is no data added to tape.\\
The CPU box represents the compute nodes with their local storage areas. The
compute nodes receive their input data only from the disk storage of the
associated site. The output data is uploaded to a disk storage area of the
site.\\
Input data that has been processed or does not fit on the hot storage is transferred
to the cloud storage. The ingress is assumed to be free of charge, while the egress
from the cloud storage to the disk storage is charged. The deletion of the data at the
cloud storage can be implemented either with an expiration time or based on a
storage limit and the popularity metric.

\section{Simulation framework}
\label{sec:simulation}
A simulation was developed to evaluate an implementation of the HCDC model.
In addition,
the simulation can generally be used to analyse different models and scenarios
combining grid and commercial cloud resources. The main feature is the
modelling of storage and network resources by simulating transfers.\\
The simulation is based on events that are scheduled to discrete time points.
An event is a subprogram that is executed at its scheduled time point during
simulation runtime. An internal clock produces the discrete time points.
The smallest time step the simulation can operate on is one second.\\
After an initialisation phase, the simulation runs an event loop. Each repetition
of the event loop handles all events of one time point. Thus, every iteration of
the event loop increases the simulation clock by the difference between the
scheduling time of the events of the current and next iteration.

\subsection{Architecture}
\label{sec:simulation_arch}
A first prototype of the simulation was implemented using Python. In favour
of control of the memory management and the performance when iterating data,
the simulation was finally developed in C++. Configuration files are formatted
in JSON.\\
\begin{figure}[t]
  \centering
  \includegraphics*[width=0.85\columnwidth]{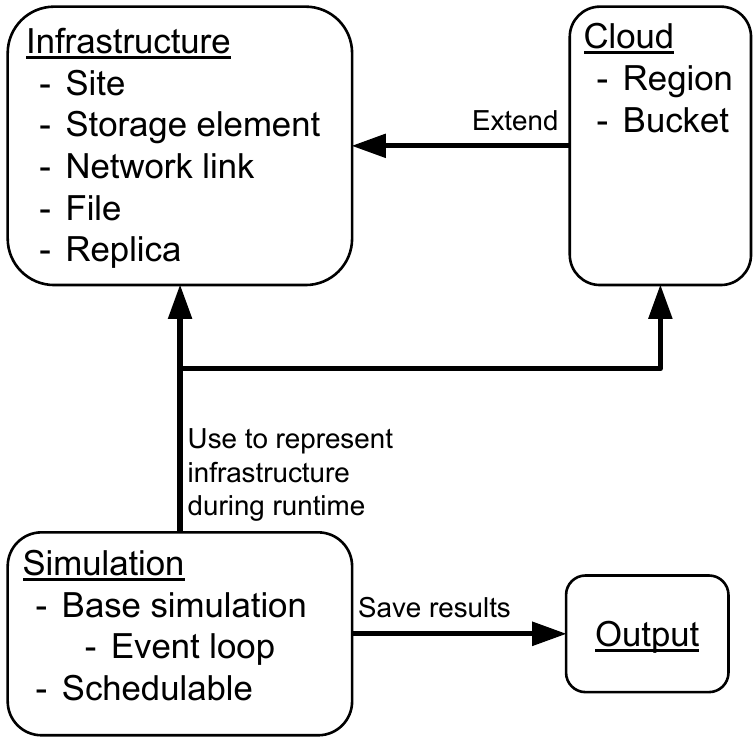}
  \caption{The four modules of the simulation. Simulation is the central
  module that manages the control flow and uses the other modules to
  create, manage, and save the simulation data.}
  \label{fig:sim_arch}
\end{figure}
The simulation consists of different modules which can be divided into four
topics as shown in Figure \ref{fig:sim_arch}.
\begin{itemize}
    \item The infrastructure module provides classes to represent the infrastructure that is simulated, e.g., storage elements, network links, files, or replicas.
    \item The cloud module extends classes of the infrastructure module to provide functionality to simulate commercial cloud resources.
    \item The simulation module provides classes that execute and control the simulation flow.
    \item The output module to persistently store data generated by the simulation.
\end{itemize}
The primary classes of the simulation module are the \texttt{BaseSimulation}
and the \texttt{Schedulable} class. Both of the two classes are designed to be
specialised in derived subclasses. The specialisation of the
\texttt{BaseSimulation}-class defines the initialisation and execution of the
simulation, including the execution of events and the management of the
simulation clock. The \texttt{Schedulable}-class serves as base class for every
event that needs to be scheduled during the simulation run. The initial events
are scheduled by the subclass instance of the \texttt{BaseSimulation}-class.
When events are executed they can reschedule themselves or create and schedule
new events depending on their implementation.\\
The infrastructure module contains data structures for all entities required
to set up the infrastructure of the model. Storage elements are objects that
address a storage area and describe its properties. They also store run time
data of the simulation, e.g., used volume and stored replicas. Each storage
element is associated to one site. Sites can contain different storage elements.
Network links represent the connection between storage elements. During run
time, they store a reference to the source and destination storage element,
as well as the traffic induced by transfers and the number of currently active
transfers. Files describe the data that can be transferred, e.g., a file object
stores the file size and the expiration time. The expiration time is the time
at that a file gets delelted. Replicas represent stored data.
During run time, a replica object contains a reference to a storage element and
a file, which means that the file is stored at the storage element.\\
Basic functionality is covered by built-in implementations in the simulation
module and can be used by customising configuration files. For example, the
HCDC model does not need a special implementation of the
\texttt{BaseSimulation}-class because the built-in implementation can set up
the required models based on corresponding configuration files.\\
A common approach to implement a new model for the simulation is to use two
configurable types of events. One type is called \emph{transfer generator}
and defines the logic of how transfers are generated. The other type is called
\emph{transfer manager} and is used to update and keep track of the states
of the generated transfers. Based on their implementation, these events
respectively create new transfers and update existing transfers. The events
reschedule themselves using a configurable time interval.\\
For all models with a transfer simulation based on bandwidth, throughput, or
transfer duration, it should be sufficient to define the parameters in a
configuration file and use one of the built-in transfer manager implementations.
In general, the transfer generator implements much more logic and thus needs a
specialised implementation for each different model.\\
There are two built-in implementations available for the transfer manager. The
first one increases the size of the destination replica of each transfer. The
amount, by which the size of the destination replica is increased, is calculated
based on the time since the last update and on the configured throughput or
bandwidth of the network link.\\
A network link can be configured in one of two modes. A configuration with a
bandwidth will equally divide the available bandwidth among the number of active
transfers. A configured throughput is not divided and is independent from the
number of active transfers. Thus, a throughput requires a reasonable distribution of
the number of transfers to deliver realistic results.\\
The other built-in transfer manager implementation updates the destination
replica based on a configured transfer duration, i.e., the transfer manager
increases the destination replica by a fixed increment each tick to ensure the
replica is complete after the configured transfer duration.

\subsection{Validation}
\label{sec:simulation_correctness}
To validate the basic functionality of the simulation, the transfers of the
ATLAS derivation input data were simulated. Since this is a process that is
already running in production, there is sufficient data available to calculate
parameters for the simulation and to provide a scale for the output.\\
The validation scenario was implemented in the simulation by creating a new
transfer generator. This transfer generator is built to only create transfers
from a configurable set of source storage elements to a configurable set of
destination storage elements. Each tick, the transfer generator processes three
steps. First, it determines the number of transfers to generate for each
source/destination pair. Second, it uniform randomly selects source files that
do not already exist at the destination. Finally, it creates the corresponding
number of transfers using the selected source and destination information.

\begin{table}[t]
   \centering
   \caption{Parameters and their configuration for the simulation validation scenario.}
   \begin{tabular}{p{0.45\columnwidth}p{0.45\columnwidth}}
        \toprule
        \textbf{Parameter} & \textbf{Value/Configuration}\\
        \midrule
        Simulated time & 59 days 19 hours\\
        Transfer mgr. update interval & 1 s\\
        Transfer gen. update interval & 10 s\\
        No. sites & 3\\
        No. initial replicas & $1000$ per site\\
        No. network links & 2 per site\\
        Throughput & 8.10 MB/s per network link\\
        File size & \begin{minipage}[t]{\textwidth}
            exponentially distributed:\\
            $\lambda = 0.61972$\\
            $10.23 \text{ MB} \leq \text{size} \leq 13.73 \text{ GB}$
        \end{minipage} \\
        No. transfers generated & \begin{minipage}[t]{\textwidth}
            exponentially distributed:\\
            $\lambda = 3.33437$
        \end{minipage} \\
        \bottomrule
   \end{tabular}
   \label{tab:correctness_params}
\end{table}

Table \ref{tab:correctness_params} lists all parameters and their configuration
that were used to simulate this scenario. The data to calculate values for
these parameters were taken from the ATLAS distributed computing monitoring
system. This monitoring system only provides the data of the past two months. For
this reason, the simulated time frame was set to almost two months as shown in the
table. Specifically, the transfer data from 2020/05/30 10:00 PM to 2020/07/29 5:00 PM
\footnote{The distributed computing infrastructure and activity was not unduly
impacted by the global COVID-19 pandemic.} of the three sites with the highest
number of transfers during this period was collected.\\
Five metrics were considered from the monitoring data. The transferred volume and
the mean transfer duration were only used as a reference for the simulation output.
The file size distribution, the number of transfers, and the transfer throughput
were also used as input parameters for the simulation.\\
The monitoring system provided the data of all five metrics only in an
aggregated form. For the file size distribution, the data was aggregated in the form
of a histogram. This histogram provided the number of files per file size. To
obtain the best data resolution, the smallest possible histogram bin
width of $128$ MB was used. For the other metrics data was aggregated in date
time histograms with a smallest possible bin width of $1$ hour.\\
The simulation parameters for the file size distribution and the number of
transfers were calculated by fitting a random distribution function to the
data. The fitting was done by maximising a log-likelihood function. Different
distributions were considered. The exponential distribution showed to
fit best for both metrics. We attribute this to the nature of the distribution
of providing a moderate average with occasional low and high values. This
property well describes both the file size distribution and the number of
transfers.\\
The throughput parameter was calculated by using the overall throughput from
the date time histogram of the $3$ observed sites. To calculate a throughput
value the mean of the histogram was taken and equally distributed by dividing
it by $3$.\\
As mentioned, the monitoring data for the number of transfers was available in
a date time histogram with $1$ hour wide bins. In reality the transfers are
not only created every hour but distributed across this hour. For this reason the
data was linearly interpolated to satisfy the $10$ seconds update interval of the
transfer generator. Furthermore, the data was uniformly distributed across the
$6$ network links because the data from the monitoring system was aggregated.
The interpolated data was used for the fit to the random distribution functions.\\
The transfer generator and transfer manager update intervals control how frequently
new transfers are created and existing transfers are updated. These parameters
can influence the resolution of the simulation. Thus, finding the optimal value
means finding a trade-off between simulation run time and simulation precision.\\
The time complexity for a single update of the transfer manager scales linearly
with the number of active transfers of all network links. From the monitoring data
it was known that the number of transfers per hour is on the order of $1000$. The
largest bin of the transfer duration histogram from the monitoring data reached
$\approx 10$ minutes. Given these values, it is not expected that the number of
active transfers reaches a scale which would noticeably increase the simulation
run time. This allowed setting the transfer manager update interval to the lowest
value of $1$ second to achieve the highest resolution.\\
The transfer generator update interval defines the minimum difference between the
creation time of transfers that were not created at the exact same time. Choosing
an excessively large update interval would result in a high but infrequent number of
transfer generations. Conversely, too small and it could result in an increase of the
simulation run time. An update interval smaller than the time between two transfer
generations does not improve the resolution. Thus, a minimum update interval of $1$
second is not reasonable for an expected rate of 1000 transfers per hour. With a
value of $10$ seconds, the overall simulation run time was on the order of
$\approx 30$ seconds.
The mean of the distribution function of the number of transfers to generate is
$1/\lambda = 1/3.33437 \approx 0.3$. This means that on average, using the $10$
seconds interval a new transfer is generated every third update.\\
At the start of the simulation, each storage element is initiated with $1000$
replicas. The transfer generator randomly selects a replica as source for each
transfer. Only files that do not have a replica at the destination and are not
in the process of being transferred to the destination can be selected.
After a completed transfer, the destination replica is deleted again to allow
transferring the replica again. The $1000$ replicas provide a sufficient pool
of selectable replicas. In case no replica meets the select conditions, a new
replica is created.

\begin{table}[t]
   \centering
   \caption{Results of simulation correctness validation. The RWD column shows the
   real world data. The Sim column shows the simulated data.}
   \begin{tabular}{lrrlr}
       \toprule
       \textbf{Metric} & \textbf{RWD} & \textbf{Sim} & \textbf{Unit} & \textbf{Diff.} \\
       \midrule
       File size & 1.74 & 1.73 & GB & 0.57 \% \\
       No. transfers & 1.77 & 1.80 & No./10s & 1.69 \% \\
       Throughput & 8.10 & 8.01 & MB/s & 1.11 \% \\
       Traffic & 3.01 & 3.11 & GB/s & 3.32 \% \\
       Transfer duration & 212.18 & 214.10 & s & 0.90 \% \\
       \bottomrule
   \end{tabular}
   \label{tab:correctness_results}
\end{table}

The evaluation was made by using the parameters from the distributions fitted
to the real world data as input and comparing the average of the output
parameters. It was not necessary to run a large number of iterations of the
simulation. Five runs have been executed and the output was compared to
verify this. The standard deviation of each observed metric were between 0\%
and 0.07\%. The standard error for the different metrics did not exceed 0.03\%.\\
Table \ref{tab:correctness_results} shows the observed metrics with real world
data values and the simulated values as well as the differences between them.
As explained before, the simulated values are the mean of five different
simulation runs. The file size in gigabytes is the mean value of the file sizes.
The number of transfers is the number of transfers that were finished
every 10 seconds. The throughput is the mean value of the throughput of all
transfers equally distributed to the three sites. These parameters are
taken from the real world data and are used as input parameters for the
simulation. Based on these parameters, the traffic and transfer durations are
computed during the simulation and are observed as output metrics. The traffic
is the summed data volume that is transferred with each transfer manager update.
Thus, the throughput metric is calculated per transfer, while the traffic is
calculated time based. The transfer duration is the mean duration of each transfer.
The highest difference between real world data and simulated data is the traffic
metric with 3.32\%.

\section{HCDC simulation}
\label{sec:simulation_hcdc}
The HCDC model features several more cases and conditions to generate transfers
than the straightforward random selection of files from the model of the previous
section. The following sections explain the implementation of the HCDC model in
the simulation software, the used parameters, and the evaluation of the results.
\subsection{Site configuration}
\label{sec:site_config}
\begin{figure}[t]
  \centering
  \includegraphics*[width=\columnwidth]{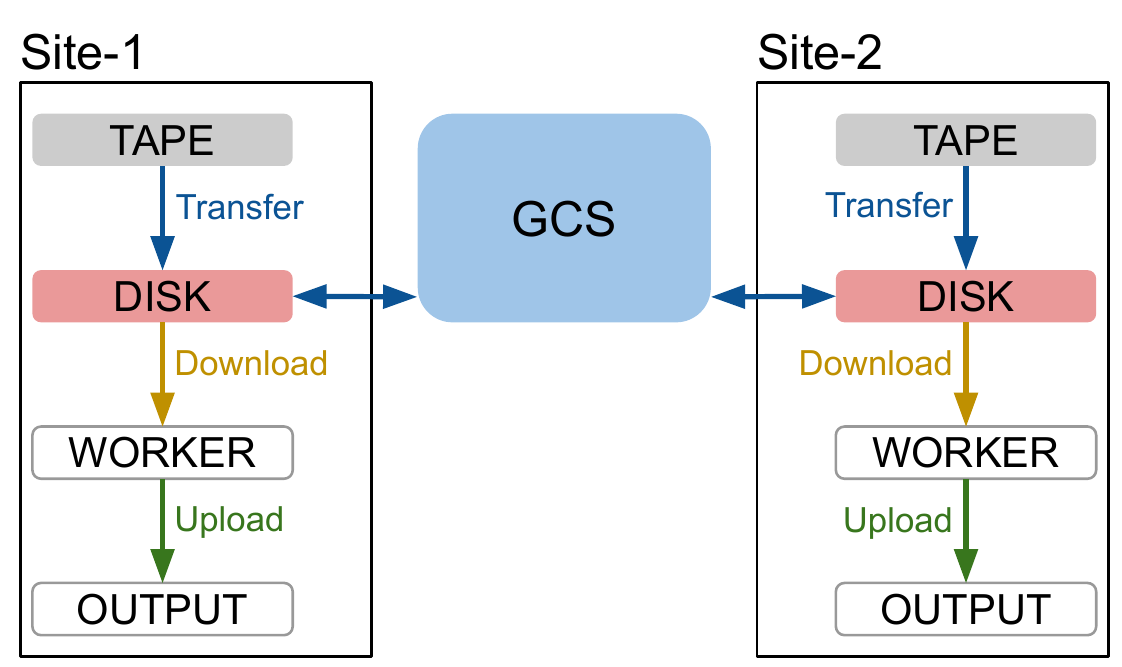}
  \caption{Implementation of the HCDC model in the simulation. It shows the
  configured storage elements for each of both sites and the network link
  setup. The network link labels name the type of transfer. GCS illustrates
  a single cloud bucket used by both sites.}
  \label{fig:model_sim_setup}
\end{figure}
Figure \ref{fig:model_sim_setup} shows the HCDC model implementation for
the simulation. It is based on the two grid sites Site-1 and Site-2 each
with four different storage elements. Network links in the simulation are
always directional. The arrows illustrate the network links and their
directions.\\
The GCS box represents a GCS bucket. In the simulated scenarios, each site
uses only the data originating from its own tape storage element although
both sites have access to all the data on GCS. To enable sites processing data
from other sites, a study would be required of how to split the workload
exactly. A possible approach would be to use a given share, e.g., $80\%$
of the jobs use local site data and $20\%$ of the jobs use remote site data.
Another approach could use the popularity metric to additionally select
remote site data.\\
The simulation was configured with the following storage element types.\\
\textbf{TAPE} storage elements provide tape storage that represents the archival
storage of the Hot/Cold Storage model. The tape storage contains one replica
of each file, and thus represents the origin of all input files. Typically,
requests to tape are kept in a queue for some time if the tape is not mounted
to optimise the reading requests. Additionally, there is a latency for
mounting, positioning, and dismounting the tape. These delays are simulated
by configuring the tape storage elements with an access latency. That means
when a queued transfer from tape storage becomes active, the start of the
actual data transferring is deferred based on the access latency.\\
\textbf{DISK} storage elements provide disk storage exclusively for input
data and represent the hot storage of the Hot/Cold Storage model. The disk
storage is used as source for the downloads of the derivation production input
files to the worker nodes. Only files on disk storage elements can be downloaded
to the worker nodes. The disk storage represents the storage area of the sliding
window in terms of the data carousel model.\\
\textbf{WORKER} storage elements are used to simulate the local storage of the
worker nodes. The simulation was designed to represent transfers with storage
and network resources so there is no default functionality for computing
resources. It is assumed that the input files always fit entirely on the
worker node. Thus, worker storage elements must not have a limit set. In general
the HCDC model can also be used without this assumption, but the simulation
implementation would require several adjustments, e.g., to support streamed
data input.\\
\textbf{OUTPUT} storage elements that represent a storage area exclusively
for output data. Output storage elements are used to store the output files of
the jobs which will be uploaded from the worker node storage.\\
\textbf{GCS} buckets are a special type of storage element used to represent
the cold storage of the Hot/Cold Storage model. The cloud module of the simulation
contains a GCS implementation. This implementation allows creating GCS buckets.
GCS buckets are storage elements that are extended by certain functionalities like
storage increase/decrease tracking, ingress/egress tracking, and cost calculation.
These functionalities implement the cost model of the cloud provider.

There are two special transfer types in this model. First, the transfers from
disk storage elements to worker storage elements are called {\em download} instead
of transfer. Second, the transfers from the worker storage element to the output
storage element are called {\em upload} instead of transfer. The difference between
a download and a transfer is that the downloaded replica is not managed by the data
management system anymore. Respectively, an upload creates a new managed replica in
the data management system. Furthermore, in the simulation downloads and uploads are
not processed by the transfer manager and they are stored in a different format in
the output module.\\

\subsection{Simulation workflow}
As explained in Section \ref{sec:simulation_arch}, a transfer generator
implements the main part of a model in the simulation. The HCDC transfer
generator is implemented and configured to simulate the continuous
derivation production. The transfer generator simulates the submission and
execution of jobs. A job has a selected input file, which is transferred
from tape to disk, downloaded to the worker storage, processed, and
subsequently deleted from the disk storage.\\
\begin{figure}[t]
  \centering
  \includegraphics*[width=\columnwidth]{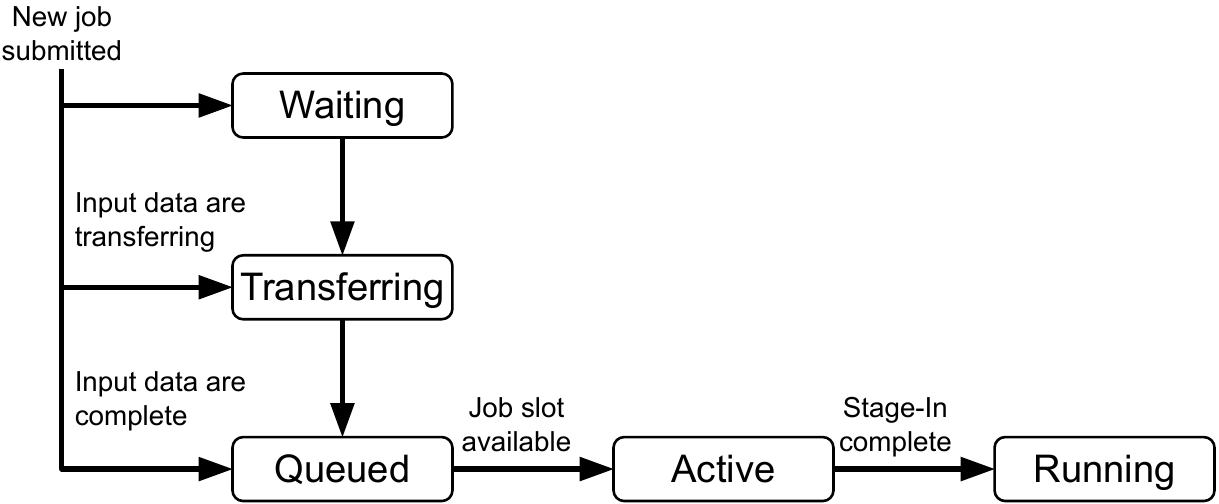}
  \caption{State transitioning of jobs during production phase.}
  \label{fig:job_statemachine}
\end{figure}
The transfer generator implements the derivation workflow by using
\textit{job objects} that transit through a state machine. Figure
\ref{fig:job_statemachine} illustrates the state machine. Each job object is
in one of the states waiting, transferring, queued, active, or running. The
arrows indicate the transition among the states. The arrow labels are the
conditions for a state transition.\\
The first operation of each transfer generator update is the submission of
new jobs. A number of new jobs, which is based on the configuration, are submitted.
For each new job, the input data is randomly selected based on the popularity.
In the following, the various states and transitions are explained in more detail.\\
\textbf{Waiting:} The needed input data is not at the disk storage element.
No transfer for the input data to the disk storage element exists. Not
enough disk storage is available to create a transfer of the required input
data to the disk storage. When disk storage becomes available, a transfer is
queued and all job objects that are waiting for these data enter the
transferring state. Job objects in the waiting state are processed in first
in, first out order.\\
\textbf{Transferring:} The needed input data is not completely available at
the disk storage element but a transfer is queued or running to replicate the
data to the disk storage element. When the transfer is complete, the state of
the job object changes to queued.\\
\textbf{Queued:} When the input data of a job is already at the disk storage
element or a transfer for the input data is completed, the job object state
transits to the queued state. In this state, the job waits for compute
resources in form of job slots to become available. The simulation can be
configured to provide a specific number of job slots per site. If a job slot
is available, the job object state changes to active.\\
\textbf{Active:} In this state, a job is occupying a job slot. A download of
the input data from the disk storage element to the worker storage element is
started. When the download is finished, the details about the download and
the job object are stored to the simulation output and the job enters the
state running.\\
\textbf{Running:} The job object simulates the derivation job execution.
Job objects in the running state are not regularly updated anymore. They
are paused for a randomly generated job execution duration, during which
time the job object is waiting for the job to finish. Subsequently,
uploads of the output data are created and the job object is deleted.

The deletion of data that is no longer required is processed at the beginning
of each transfer generator update. Obsolete replicas on the disk storage element
that already have a replica on the GCS bucket are deleted immediately.
The data that is not available on the GCS bucket is migrated there, i.e.,
transferred from the disk storage element to the GCS bucket and subsequently
deleted on the disk storage element.

\subsection{Parameters}
\label{sec:hcdc_params}
\begin{table}[t]
   \centering
   \caption{Parameters and their configuration for the simulation of
   the HCDC model.}
   \begin{tabular}{p{0.45\columnwidth}p{0.45\columnwidth}}
        \toprule
        \textbf{Parameter} & \textbf{Value/Configuration}\\
        \midrule
        Simulated time & 90 days\\
        Transfer mgr. update interval & 1 s\\
        Transfer gen. update interval & 10 s\\
        No. sites & 2\\
        No. initial replicas & $10^6$ per site\\
        Popularity &
        \begin{minipage}[t]{\textwidth}
            geometrically distributed:\\
            $p = 0.1$\\
            $1 \leq x < 50$
        \end{minipage} \\
        Input file size &
        \begin{minipage}[t]{\textwidth}
            exponentially distributed:\\
            $\lambda = 0.026$\\
            $9.76 \text{ MB} \leq \text{size} \leq 134 \text{ GB}$
        \end{minipage} \\
        No. jobs submitted &
        \begin{minipage}[t]{\textwidth}
            normally distributed:\\
            $\mu = 0.63366$\\
            $\sigma = 0.37292$\\
            $n \geq 0$
        \end{minipage} \\
        Job duration &
        \begin{minipage}[t]{\textwidth}
            exponentially distributed:\\
            $\lambda = 0.00409$\\
            $t \geq 16.666$ minutes
        \end{minipage} \\
        \bottomrule
   \end{tabular}
   \label{tab:hcdc_params}
\end{table}
Table \ref{tab:hcdc_params} shows the general and site specific parameter
values used for the HCDC model. The monitoring data was taken from different
sources for the HCDC model. Data to calculate the throughput was taken from
the transfer monitoring data. The other parameters such as input file size
distribution was taken from the job monitoring data. Contrary to the transfer
monitoring data, the job monitoring data is not limited to the past two months.
However, since the mean throughput is calculated from the data, it is assumed
that the mean value is also representative for the time of three months. The
monitoring data for two months was taken for the time from 2020/07/08 12:00 PM
to 2020/09/06 12:00 PM. The monitoring data for three months was taken for the
time from 2020/06/08 12:00 PM to 2020/09/06 12:00 PM.

First real world tests with the HCDC model would be limited to rather small
scale deployments, especially in terms of the number of sites. This prevents
a large wasting of resources in case of issues with the implementation.
Additionally, it allows acquiring an impression of the model without incurring
excessively large cloud costs. The monitoring data showed that in total $80$ sites
processed $\approx 6.5$ million derivation jobs during the observed $3$ months.
The $2$ sites with the most number of derivation jobs each processed $\approx 0.5$
million jobs. To be able to use the same job submission configuration and to keep
the first evaluation of the model clear, only these $2$ sites were simulated. The
simulated time was set to $90$ days according to the monitoring data.\\
The required real time to simulate one day was typically below one second on
an average virtual machine. Thus, a value of $1$ second could be used for the
transfer manager update interval to achieve the best possible resolution. The
vast majority of the simulation run time is spent in the transfer generator. 
The transfer generator update interval was chosen for the same reason as in
Section \ref{sec:simulation_correctness} but based on the number of submitted
jobs.\\
The file size distribution was calculated using the same approach as in
Section \ref{sec:simulation_correctness}. The used monitoring data did
not provide the size of each input file but only the total input volume
of each job. Using this monitoring data makes the assumption that one
transfer from the tape storage provides data for at least one job. That
implies a reasonable data placement on tape and that the workflow
management system is able to structure jobs to transfer data bunches
from tape and process them. These implications are still challenging
objectives. To improve their impact on the accuracy of the simulation,
a more detailed model of the job submission would be required.
Furthermore, a more detailed tape model including the used data placement
strategies would be required.\\
Since one file corresponds to the input data of one job, the number of
initial replicas can be based on the number of jobs. The number of
finished jobs in the monitoring system was $\approx 10^6$. Thus, an
equal number of files was created.\\
As mentioned before, the number of times data was processed is used as
the popularity metric. This information can be collected from the central
production database. Most data was processed once, exponentially falling-off
to $50$ times processed. A geometrically distributed random function can
approximate this falloff with the chosen parameters of $p = 0.1$ and the
limits of $1 \leq x < 50$.\\
The number of jobs to submit is generated by a normally distributed random
function. The duration of each job is generated by an exponentially
distributed random function. The estimation of the parameters of these
functions follows the same approach as the number of transfer generation
in Section \ref{sec:simulation_correctness}. Since the number of jobs to
submit is fitted to the monitoring data, no job slot limitation is
configured.\\
No specific configuration for the number of output files and volume of the
output data was used. This is because in this model these metrics do not
affect the bandwidth or cost. Only the job slot is blocked until all
uploads of the output files are finished. Furthermore the metrics depend
directly on the number of jobs finished, and thus could be calculated after
the simulation.

\begin{table}[t]
   \centering
   \caption{Network configuration of the simulated model.}
   \begin{tabular}{p{0.1\columnwidth}p{0.1\columnwidth}p{0.1\columnwidth}rr}
        \toprule
        \textbf{Site} & \textbf{Source} & \textbf{Dest.} & \textbf{Max. active} & \textbf{Value}\\
        \midrule
        Both & GCS & Disk & 100 & 294.00 MB/s\\
        Both & Disk & GCS  & 100 & 500.00 MB/s\\
        Both & Disk & Worker & N/A & 88.24 MB/s\\
        Site-1 & Tape & Disk & 100 & 22.62 MB/s\\
        Site-2 & Tape & Disk & 100 & 62.35 MB/s\\
        \bottomrule
   \end{tabular}
   \label{tab:hcdc_networkcfg}
\end{table}
The other parameters define the network configuration. Table
\ref{tab:hcdc_networkcfg} shows the used values. The first two rows show
the throughput for the network links between the GCS bucket and the disk
storage elements. The third row shows the throughput for downloads. These
links are configured equally for Site-1 and Site-2. For transfers managed
by the transfer service, the maximum number of active transfers was
limited to $100$ according to the real world limits. The number of
downloads is not explicitly limited.\\
For the throughput parameters, the mean value of the monitoring data was
taken. Compared to the transfers between the disk storage element and the
GCS bucket, the download throughput seems to be low, but there is no
limitation on the number of active downloads. As explained in Section
\ref{sec:simulation_arch}, values configured as throughput are not divided
among the number of active transfers.\\
For the throughput from and to the GCS bucket, only monitoring data from
small scale manual tests was available. Because of the small scale, these
values might contain uncertainties and must be adjusted when larger scale
data is available.\\
Another configuration parameter to consider is the access latency of the
tape storage elements. Creating a proper model to estimate the access
latency would be a complex topic itself and would require detailed log
data from real tape systems. For this reason, a constant average value of
$30$ minutes was used for the access latency. In addition, tests were
done using normally distributed random values for each transfer.

The pricing information of the commercial cloud storage is defined in
a configuration file. It contains the pricing details of different
storage categories, different grades of cost depending on the stored
volume, and the network cost depending on the egress destination. The
file was created based on the public pricing data from the GCP
documentation on the 2020/09/10. For the simulation, the standard
storage class for a regional bucket was used.\\
ATLAS and the VR Observatory worked on research and development projects
in collaboration
with Google. During this work special prices were negotiated.
Furthermore, Google supports different peering methods than using the
internet. Connection to GCP using a non public network could immensely
reduce the network cost. For example, the price for downloading to the
internet in Europe is between $0.08$ and $0.12$ USD/GiB. Using the
direct peering option the cost are reduced to $0.05$ USD/GiB in Europe.
The interconnect peering option charges only $0.02$ USD/GiB \cite{gcp}.
These peering methods typically require a physical network connection
to an internet exchange point.

\begin{table}[t]
   \centering
   \caption{Different storage limits per configuration.}
   \begin{tabular}{lp{0.25\columnwidth}p{0.23\columnwidth}p{0.23\columnwidth}}
        \toprule
        \textbf{Cfg.} & \textbf{Disk limit} & \textbf{GCS limit} & \textbf{Tape limit}\\
        \midrule
        \emph{I} & N/A & 0 & N/A\\
        \emph{II} & 100 TB & 0  & N/A\\
        \emph{III} & 100 TB & N/A & N/A\\
        \bottomrule
   \end{tabular}
   \label{tab:hcdc_storagelimits}
\end{table}
Three different configurations of the HCDC model have been simulated.
All configurations use the same job submission, file, and network
parameters. Since these parameter are fitted to real world data, the
limits are known to be achievable by the real world system. The
differences between the configurations are the storage limits of the
disk storage elements and the GCS bucket. Table
\ref{tab:hcdc_storagelimits} shows the storage limits per configuration.\\
\textbf{\emph{Configuration I}} has the limit of the GCS bucket set
to zero to prevent the usage of the GCS bucket. In addition, no
limit is set on the disk storage elements. With this configuration,
all input data is transferred from the tape storage element to the
disk storage element and is kept at the disk storage element. The
simulation results should be comparable to the current ATLAS
derivation production workflow since the only difference is the
initial transfer from the tape storage element to the disk storage
element.\\
\textbf{\emph{Configuration II}} has the same limit value of zero
on the GCS bucket. In addition, a limit of $100$ TB is set on each
disk storage element. This configuration of the model shows how the
results would be if there was no cloud storage to cache the input
data. In this scenario, the input data must be transferred from the
tape storage element to the disk storage each time the data is
required.\\
\textbf{\emph{Configuration III}} is without a limit on the GCS bucket,
but with the limit of $100$ TB on the disk storage elements. In this
way, the fully combined model is simulated with all storage areas
usable. This configuration allows analysing the difference when adding
the cloud storage as cache.

\subsection{Evaluation}
\label{sec:evaluation}
The first metric that was evaluated is the number of finished jobs. From the
data of the monitoring system, it is known that the number of finished jobs
should be $\approx 10^6$. As explained in Section \ref{sec:hcdc_params},
it is expected that the results of \emph{configuration I} are similar to the
real world data. In \emph{configuration II}, the data on the disk storage element
is deleted after it has been processed because of the limit of disk storage.
In addition, the GCS bucket is not used. This leads to transferring the data
from the tape storage element to the disk storage element each time the data
is required. Compared to \emph{configuration I} this results in an increase
of total required transfers. With the increased number of transfers and the
additional access latency for tape access, \emph{configuration II} is expected
to show a decrease in the number of finished jobs.\\
\begin{table}[t]
   \centering
   \caption{Mean number of finished jobs and mean volume downloaded with their
   standard deviation for \emph{configuration I, II,} and \emph{III} of $20$ simulation runs.
   The number in the brackets is the corresponding standard error (SE).}
   \begin{tabular}{lrr}
       \toprule
       \textbf{Cfg.} & \textbf{No. jobs done (SE)} & \textbf{Download volume (SE)} \\
       \midrule
       \emph{I}   & 996k $\pm$ 0.05\% (0.01\%) & 41.11 PB $\pm$ 0.20\% (0.04\%)\\
       \emph{II}  & 853k $\pm$ 0.11\% (0.02\%) & 35.28 PB $\pm$ 0.24\% (0.05\%)\\
       \emph{III} & 996k $\pm$ 0.05\% (0.01\%) & 41.02 PB $\pm$ 0.38\% (0.08\%)\\
       \bottomrule
   \end{tabular}
   \label{tab:hcdc_results}
\end{table}
Table \ref{tab:hcdc_results} shows the number of finished jobs and the total
volume downloaded for \emph{configuration I, II,} and \emph{III}. Comparing the
results of \emph{configuration I} and \emph{II} already gives an impression of
the expected impact of the limit on disk storage. The limit results in
$\approx 15\%$ fewer jobs finished and in $\approx 14\%$ less input volume
downloaded.\\
It is conceivable that the difference between the number of finished jobs of
\emph{configuration I} and of \emph{configuration II} increases as the simulated
time frame increases. The explanation is that in the beginning both configurations
behave similar because all the data is on tape storage. At some point in
\emph{configuration I} all the data will be available on disk storage. Thus, the
data can be directly downloaded from the disk storage without transferring it
from tape first. This is not the case for \emph{configuration II} because of the
disk storage limit. With a limited disk storage, further jobs can only be started
if sufficient disk storage space is made available by the running jobs.\\
\emph{Configuration III} uses the same values as \emph{configuration II}
except that the GCS bucket has no limit. As Table \ref{tab:hcdc_results}
shows, the results are almost equal to \emph{configuration I} in terms of
number of jobs finished and volume downloaded. That means in terms of these
metrics the cloud storage is able to compensate the limit of the disk storage
element.

\begin{figure}[t]
  \centering
  \includegraphics*[width=\columnwidth]{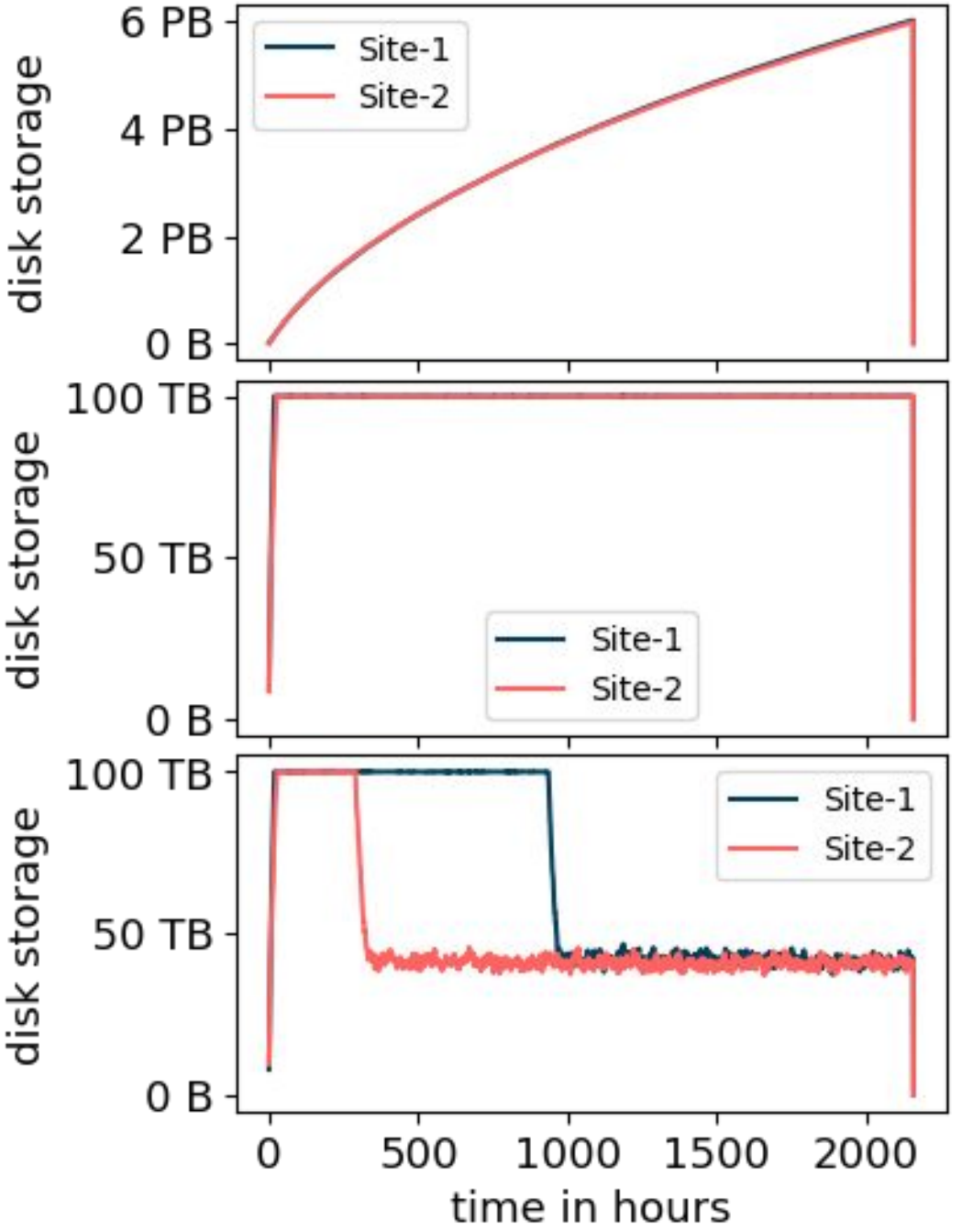}
  \caption{Increase of used storage of the disk storage element for
  \emph{configuration I} (top), \emph{II} (middle), and \emph{III} (bottom).
  The used storage increase for Site-1 and Site-2 overlap because they are
  very similar for \emph{configuration I} and \emph{II}.}
  \label{fig:storage_filling}
\end{figure}
Figure \ref{fig:storage_filling} shows the increase of used storage over time
for each configuration and each disk storage element. In \emph{configuration I}, the
disk storage element is unlimited. This results in a quick increase of used
storage at the beginning. The more data is transferred to the disk storage, the
more the increase flattens out.\\
The disk storage of \emph{configuration II} is limited to $100$ TB as reflected
by the corresponding graph in Figure \ref{fig:storage_filling}. The used storage
is fluctuating slightly below $100$ TB because of the continuous deletion of old
and creation of new replicas.\\
At the beginning of the simulation, \emph{configuration III} is equal to
\emph{configuration II} because all the data is at the tape storage element and
is required to be transferred to the disk storage element prior to its
processing. During this time, more jobs are submitted on average than can
be processed because of the tape throughput and access latency. This results
in a backlog of jobs waiting for disk storage. At some point, a sufficient
amount of data is stored on GCS. After this point, less jobs are submitted
on average than can be processed. This results in a quick processing of the
backlog of submitted jobs and afterwards a reduced storage requirement.
Site-2 requires less time to reach this point because the tape throughput is
larger than at Site-1.\\
As mentioned before, additional tests with a random normally distributed
tape access latency were made. The access latency was in the range between
$0$ and $90$ minutes. The first runs were made with a mean value of $30$
minutes and a standard deviation of $10$ minutes. The number of finished
jobs and the volume transferred, did not change significantly for any
configuration. For \emph{configuration II}, an increased standard deviation
slightly reduced the number of jobs because at some point the transfers with
large access latency dominate and create a transfer queue backlog. The mean
value of the random distribution has a noticeably stronger impact. Increasing
the mean to $60$ minutes +- 15 minutes reduced the number of finished jobs
by $\approx 20$\% for \emph{configuration II}. The number of finished jobs
for \emph{configuration I} and \emph{III} were reduced by $2$\% and $4$\%,
respectively.

Approximately the same number of jobs are submitted for each configuration but
in the case of \emph{configuration II} $\approx 15\%$ fewer jobs are finished.
This leads to the conclusion that some jobs spent more time in early states
of the job object state machine. Each job object stores different time points
from its submission to its deletion. The \emph{job waiting time} is the time
span from the submission of a job until the job is queued. This includes the
time the job must wait for disk storage, the time that the corresponding
transfer spends in the queue, and the time to finish the transfer.\\
\begin{figure}[t]
  \centering
  \includegraphics*[width=\columnwidth]{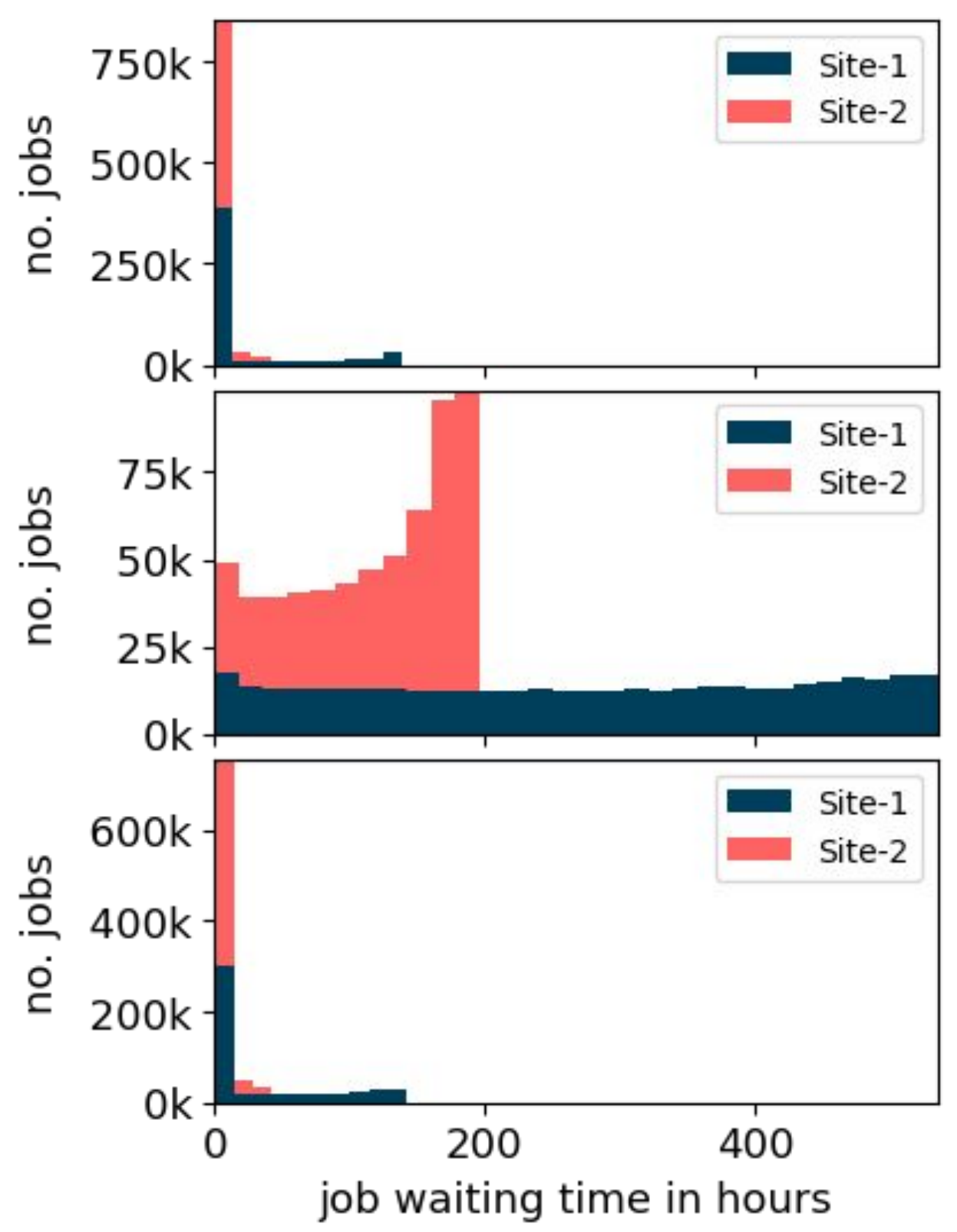}
  \caption{Job waiting time distribution of the HCDC model for \emph{configuration I} (top),
  \emph{II} (middle), and \emph{III} (bottom). The number of bins in the top and bottom
  histogram was reduced from $30$ to $10$ to improve visibility of the first bin.
  The bins are stacked.}
  \label{fig:wtfd_all}
\end{figure}
Figure \ref{fig:wtfd_all} shows histograms for the job waiting time for each
configuration. The top histogram shows the data from \emph{configuration I}.
As expected, the vast majority of jobs have a small job waiting time because
the disk storage is not limited and the data is transferred only once from
the tape storage element to the disk storage element. At the beginning of the
simulation, a small backlog of transfers occurs because of the access latency
of the tape storage and because the network links from the tape storage allow
only $100$ active transfers at the same time. The transfers at the end of this
backlog represent the outliers in the histogram with a job waiting time
of $\approx 150$ hours.\\
The second histogram shows the data from \emph{configuration II}. The histogram
shows that more jobs have a larger job waiting time when disk storage is
limited. Furthermore, the job waiting time distribution is different between
Site-1 and Site-2. Similar to \emph{configuration I}, a backlog of waiting
jobs develops in \emph{configuration II}. The backlog becomes larger than in
\emph{configuration I} because the data must be frequently transferred from tape.
The jobs at the end of the backlog contribute to the last bins of the histogram.
The jobs whose data is already on disk storage contribute to the bins at the
front of the histogram. The jobs whose data transfers are already queued make
the bins in the middle of the histogram. The backlog of Site-1 is larger than
of Site-2 because of the smaller tape throughput at Site-1. The histogram
shows that the job waiting time of Site-1 is distributed $\approx 2.5$ times
larger than of Site-2.\\
The third histogram shows the data from \emph{configuration III}. The result is
similar to the histogram of \emph{configuration I}, but it contains more jobs
with durations above $0$ hours. This can be explained by the inclusion of the
transfer duration into the waiting time. In this case the transfer time from
the GCS bucket to the disk storage element is added.

\begin{figure}[t]
  \centering
  \includegraphics*[width=\columnwidth]{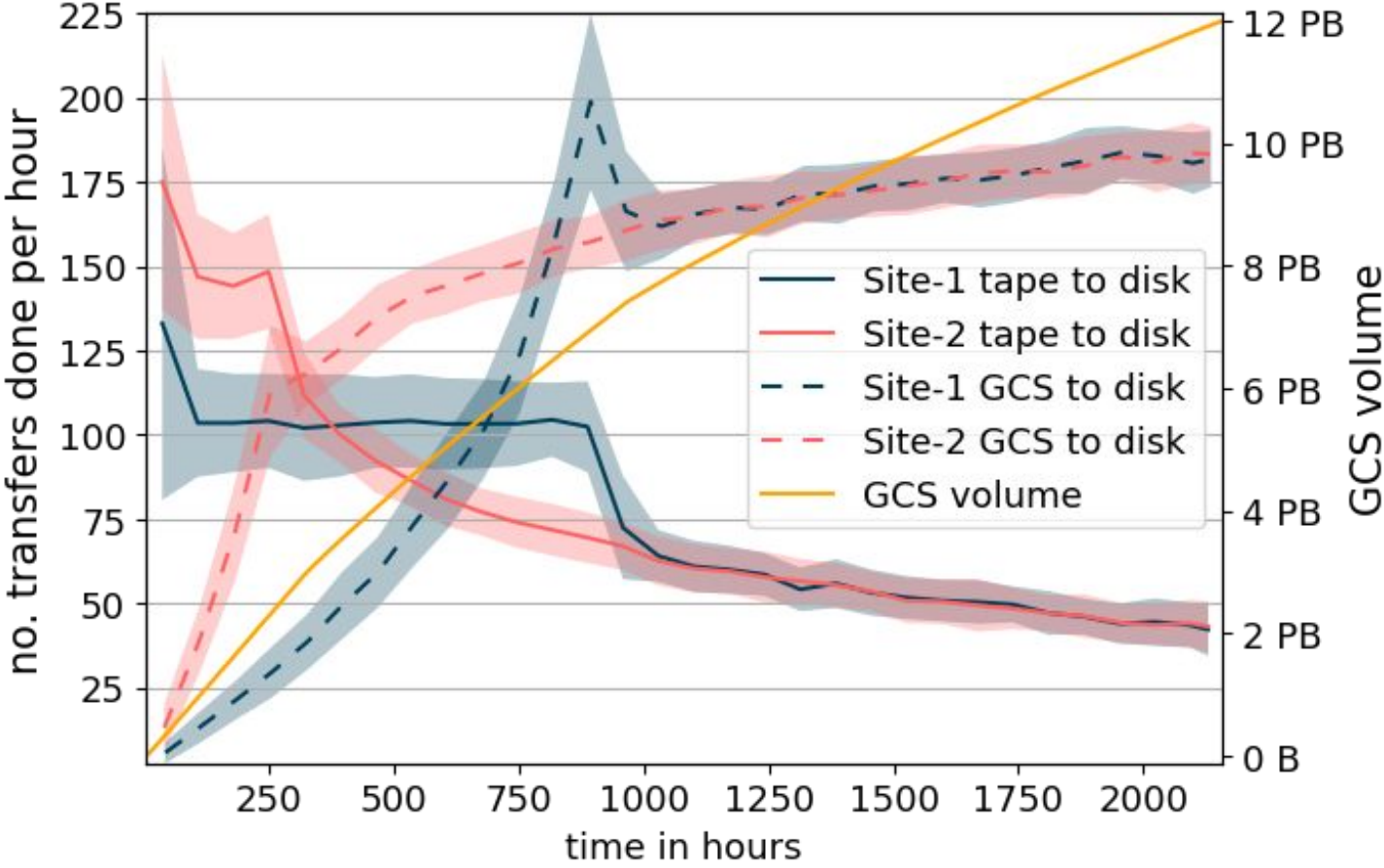}
  \caption{The solid blue and red line show the number of transfers from tape to
  disk per hour for each site. The dashed lines show the number of transfers from
  GCS to disk per hour for each site. The orange line shows the used GCS volume.}
  \label{fig:cold_fill}
\end{figure}
Figure \ref{fig:cold_fill} illustrates the usage of GCS. This graphic
is only available for \emph{configuration III} because it is the only
configuration using GCS. Data for the blue and red lines is aggregated
in the form of count per hour. To reduce the fluctuations of these aggregated
values, a mean filter was applied. The corresponding standard deviation is
shown by the line contours.\\
The blue and red lines show the number of transfers for Site-1 and Site-2
respectively. The solid versions of these lines show the hourly number of
transfers from tape to disk, while the dashed versions show the hourly
number of transfers from GCS to disk. The orange line shows the used volume
of GCS.\\
Since all the data is solely on tape storage at the beginning, the number of
transfers from GCS to disk and the used GCS volume start at $0$. The blue
and red solid lines show that the most data is transferred from tape to disk
storage at the beginning.\\
All the data that is transferred from the tape to the disk storage is
subsequently transferred from the disk storage to GCS. Furthermore,
this implementation of the model does not delete the data at GCS. That
means the orange line increases dependent on the dashed lines. At some
point the most popular data is stored at GCS. This is when the dashed
line exceeds the corresponding solid line. That means more data can be
transferred from GCS to the disk storage than from the tape storage to
the disk storage. The increase of the stored volume at GCS flattens as
the most popular data is replicated to GCS.\\
The Figure also indicates the backlog effect explained earlier. At the
beginning more jobs are submitted than can be processed because the data
comes from tape. When the most popular data is available at GCS, the
backlog can be processed. This effect is the reason for the peak of the
dashed blue line and the drop afterwards. This effect is also observable
for Site-2 but because of the larger throughput it occurs earlier and
is smaller.

\begin{table}[t]
   \centering
   \caption{Mean and standard deviation of transferred volume between storage
   elements for \emph{configuration I, II,} and \emph{III} of $20$ simulation runs.}
   \begin{tabular}{lllr}
       \toprule
       \textbf{Cfg.} & \textbf{Site} & \textbf{Transfer} & \textbf{Volume (SE)} \\
       \midrule
       \emph{I}   & Site-1 & tape to disk &  6.75 PB $\pm$ 0.28\% (0.06\%)\\
       \emph{I}   & Site-2 & tape to disk &  6.74 PB $\pm$ 0.29\% (0.06\%)\\
       \emph{II}  & Site-1 & tape to disk &  8.85 PB $\pm$ 0.07\% (0.04\%)\\
       \emph{II}  & Site-2 & tape to disk & 13.04 PB $\pm$ 0.20\% (0.01\%)\\
       \emph{III} & Site-1 & tape to disk &  6.74 PB $\pm$ 0.30\% (0.07\%)\\
       \emph{III} & Site-2 & tape to disk &  6.75 PB $\pm$ 0.19\% (0.04\%)\\
       \emph{III} & GCS    & GCS to disk  & 24.99 PB $\pm$ 0.46\% (0.10\%)\\
       \bottomrule
   \end{tabular}
   \label{tab:hcdc_transfer_results}
\end{table}
Table \ref{tab:hcdc_transfer_results} shows detailed numbers of the transfer
statistics. The transferred volume of Site-1 and Site-2 for \emph{configuration I}
is almost the same. This is because at some point the most popular data is
available at the disk storage and does not need to be transferred again from
the tape storage. About $13.5$ PB of data was transferred to the disk storage.
Comparing this amount to the volume downloaded from disk to worker storage of
$41.11$ PB from Table \ref{tab:hcdc_results} underlines the conclusion that
data is reused.\\
The transferred volume for \emph{configuration II} shows that much more data is
required to be transferred from tape storage. This is because the data at the
disk storage is deleted after processing and have to be re-transferred in case
it is required again. The difference of the transferred volume between Site-1
and Site-2 makes clear that the tape-to-disk throughput is the bottleneck.\\
The transferred volume from tape to disk storage for \emph{configuration III}
shows similar numbers to \emph{configuration I}. The volume of $6.75$ PB seems
to be the required storage for the most popular data. Once this data is available
at the disk storage for \emph{configuration I} or at the GCS for
\emph{configuration III}, less tape-to-disk throughput is required.\\
The volume transferred from GCS to disk is split between both sites, which makes
a mean of $\approx 1.6$ GB/s per site in 3 months. As mentioned before, the
configured throughput for the network links is based on small scale real world
tests. This configuration must be adjusted when increasing the scale of the
simulation in order to achieve realistic results. The real world tests of the
VR Observatory \cite{vrogcp} resulted in a similar throughput estimation. Using a
simple regression over 4 data points, they calculated a bandwidth of
$\approx 1.5$ GB/s without special performance tuning.\\
At the end of the simulation there is $\approx 12$ PB of data stored at GCS.
$\approx 6.8$ PB have not been transferred out of GCS during the simulated
time. The number of times a file was recalled from GCS is in the range from
$0$ to $45$. The files that were recalled less than 25 times are responsible
for more than 90\% of the traffic from GCS to the site disk storage.

\begin{table}[t]
  \centering
  \caption{Mean GCS costs for each month with the standard deviation and
  standard error for $20$ simulation runs. Pricing information were taken
  in US Dollar (USD) from the GCP documentation at 2020/09/10.}
   \begin{tabular}{lp{0.35\columnwidth}p{0.37\columnwidth}}
      \toprule
      \textbf{Month} & \textbf{Storage cost in USD (SE)} & \textbf{Network cost in USD (SE)} \\
      \midrule
      1 &  82k $\pm$ 0.10\% (0.02\%) & 330k $\pm$ 0.39\% (0.08\%)\\
      2 & 211k $\pm$ 0.16\% (0.03\%) & 729k $\pm$ 0.31\% (0.07\%)\\
      3 & 293k $\pm$ 0.23\% (0.05\%) & 807k $\pm$ 0.25\% (0.05\%)\\
      \bottomrule
  \end{tabular}
  \label{tab:gcs_cost}
\end{table}
Table \ref{tab:gcs_cost} lists the mean cost of the cloud resources per
month used by the simulated model. The storage cost increase from the
first month on because at the beginning the data is solely on tape and is
gradually transferred to the cloud storage. With an increasing amount of
data at GCS, more data can be transferred from GCS to the disk storage. Thus,
the network cost are increasing from month to month. The network cost are
in general larger than the storage cost since the price for network traffic
is typically much larger than the storage price.

\section{Conclusion}
\label{sec:conclusion}
This contribution describes the data carousel model and the Hot/Cold Storage model.
Currently, both models are being developed and evaluated by the ATLAS collaboration.
Specific workflows can benefit from a combination of these two models into the HCDC
model. The HCDC model can be implemented in different variations, e.g., using
different storage types to implement the Hot/Cold Storage model part or improving
the impact of the data carousel model part by adjusting the workflow.\\
To evaluate variations of the HCDC model and estimate certain metrics, a simulation
was developed. The simulation was developed as a multi-purpose software framework.
This framework can be used to simulate various kinds of models related to distributed
computing and commercial cloud storage and network resources.\\
On an average virtual machine, the validation model requires less than a second in
real time to simulate one day. The more complex HCDC model required less than two
seconds in real time for one simulated day. The memory consumption mainly depends
on the number of files and replicas. Both types require $68$ bytes for each object
instance. The memory consumption of the HCDC simulation starts at $\approx 480$ MB
and peaks to $\approx 500$ MB during run time.\\
To show the correctness of the basic functionality of the simulation framework,
an existing workflow with sufficient monitoring data was simulated and evaluated.
The used workflow is the transfer of ATLAS derivation input data between sites.\\
Section \ref{sec:evaluation} discussed the evaluation of the simulation of the
HCDC model. The HCDC model implementation used GCS for cold storage and assumed
a continuous production of derivation data. As shown in the evaluation, using the
HCDC model in combination with GCS allows reducing disk storage requirements while
maintaining the input data throughput and job throughput. However, the GCS and
network usage induce additional cost. Whether these costs are worth the additional
throughput has to be decided for each specific case by the collaboration.\\
A simulation tool, such as the presented simulation, can assist by calculating
the parameters for the decision whether to use cloud resources. Assuming the HCDC
model, parameters can be considered as fixed or variable. Fixed parameters are
dictated by the problem description and the existing resources, e.g., bandwidths,
input volume to process, number of jobs to run, or the popularity of files.
Variable parameters can be changed directly or change indirectly in dependence
on other variable parameters. For example, the cost and the job throughput
change in dependence on a potential GCS limit. The time limit to process all the
data change in dependence on the job throughput. The required GCS limit depends
on the available disk storage limit. Given a specific use case with well defined
limits of the variable parameters, the simulation can be used to estimate
the optimal balance among the parameters.\\
A typical consideration is to compare the cost induced by the cloud provider
against the benefits of the additional resources. In Section \ref{sec:evaluation},
the benefits of additional resources were measured in form of number of jobs
finished resulting from the input volume throughput. From these values,
more specific metrics can be derived, e.g., job slot saturation or the volume
of output data. For example, \emph{configuration II} finished $\approx 15$ \%
less jobs but \emph{configuration I} required $\approx 12$ PB more disk storage
and \emph{configuration III} induced more than $2$ million USD of cost for
three months.

There are several topics that should be prioritised in future work. These
topics are split in two parts. First, the used parameters and models must
be improved. Second, the presented implementation misses some concepts that
are required for more realistic simulations. When these topics have been
addressed, a specification could be defined for the variable parameters
and the simulation can be used to search the optimal values. These values
can be used for real world tests to analyse the accuracy of the estimation
of the simulation.\\
The first part of the primary future work addresses that the used
parameters are very specifically fitted to the monitoring data. The
parameters are realistic as long as the same monitoring data is used.
However, changing one of the parameters might invalidate the other
parameters. For this reason, the following improvements should be made. 
\begin{enumerate*}[label=(\roman*)]
    \item As described in \ref{sec:hcdc_params}, the job submission model
    should be adjusted, e.g., based on a job slot limit per site.
    \item The number of input files for each job must be generated based
    on a more realistic distribution.
    \item The network links should be configured with a shared bandwidth,
    which requires more detailed information about the real network
    topology. In addition, a background traffic should be added to shared
    bandwidth links assuming the links are not exclusively used for the
    simulated scenario.
    \item The throughput between GCS and the WLCG was based on rather small
    scale tests and need to be tested with larger transfer volumes or modelled
    differently.
    \item The simulated HCDC model used a statically assigned popularity
    based on a fitted random function. This could be replaced by a dynamical
    assignment. A straightforward approach would be the least recently used
    concept, which is a established CPU caching technique. More advanced
    approaches could implement models of existing research about popularity
    based data replication \cite{cms_caching, atlas_caching}.
    \item The tape access latency is a strong simplification of real tape
    systems. A more realistic model based on logs of a real tape system
    would be required.
    \item The effect of increasing the simulation time has to be investigated
    in future work. The simulation scales to much larger times, e.g., one year.
    However, this requires to implement the improvements of the parameters
    because the current parameters might not be valid for more than 3 months.
\end{enumerate*}\\
The second part of the primary future work is the implementation of
additional concepts. There are two concepts that should be added to
the simulation. Currently, it is possible to limit the GCS but there is no
mechanic that deletes data on GCS. This is an essential feature to be able
to define narrower limits on GCS and subsequently reduce storage cost.
Different deletion strategies are possible, e.g., setting a storage capacity
threshold. After surpassing this threshold, the data is deleted from GCS based
on the popularity.\\
The other concept that should be implemented is
transfers between sites. Using WLCG resources, the data to process is
typically distributed among various sites to benefit from different
resources and optimally distribute the workload. The challenge of
implementing this into the simulation is the creation of a realistic
model that specifies the amount and the selection of data to transfer
between sites.\\
After the primary future work topics have been implemented, further
adjustments to the HCDC model are possible. For example, the GCS
egress cost are even more critical than the GCS cost. These could be
reduced by improving the deletion at the disk storage element and making
the caching strategy more intelligent. Moreover, the utilisation of the
tape storage decreases as the GCS usage increases. Instead of always
preferring GCS over the tape storage, both storage categories could be
used optimally. One approach for this would be to store the largest files
only on the tape storage and not to transfer them to GCS. The access to
the tape storage becomes more performant for larger files. The tape
storage usage would be improved and the egress cost of the cloud storage
would be further reduced.\\
Another approach to potentially reduce egress cost is to utilise cloud
compute resources and derive data directly inside the cloud. Although
this approach adds cost for the compute resources, it also reduces the
egress cost because only the smaller derived data have to be transferred
out of the cloud.\\
Further, future work could also investigate the HCDC model assuming the derivation
production workflows being organised in campaigns. As explained in Section
\ref{sec:hcdc}, in this case it is expected that the Hot/Cold Storage model
part becomes less impactful but the data carousel model part can be used
optimally.

\section*{Acknowledgements}
This work was done as part of the distributed computing research and development
programme of the ATLAS Collaboration, and we thank the collaboration for its
support and cooperation. We are grateful to the many ATLAS Distributed Computing
teams, especially Johannes Elmsheuser and Alexei Klimentov for fruitful discussions
and providing feedback on this work. Furthermore, frequent meetings with experts
from Google were enormously helpful and they are much appreciated. Also special
thanks go to Christian Albrecht for his valuable comments and support.

\section*{Conflict of interest}
On behalf of all authors, the corresponding author states that there is no conflict of interest.

\bibliographystyle{unsrt}

\begin{thebibliography}{10}

\bibitem{atlas:paper}
{ATLAS} {Collaboration}.
\newblock The {ATLAS} {Experiment} at the {CERN} {Large} {Hadron} {Collider}.
\newblock {\em JINST}, 3:S08003, 2008.

\bibitem{cms}
{CMS} {Collaboration}.
\newblock {The CMS Experiment at the CERN LHC}.
\newblock {\em JINST}, 3:S08004, 2008.

\bibitem{LSST}
Ivezi{\'{c}} Z, Kahn SM, Tyson JA, et~al.
\newblock {LSST: From Science Drivers to Reference Design and Anticipated Data
  Products}.
\newblock {\em The Astrophysical Journal}, 873(2):111, 2019.

\bibitem{ska}
Weltman A, Bull P, Camera S, et~al.
\newblock {Fundamental Physics with the Square Kilometer Array}.
\newblock {\em Publications of the Astronomical Society of Australia}, 37,
  2020.

\bibitem{GRIDHEP04}
Avery P.
\newblock {Grid Computing in High Energy Physics}.
\newblock {\em AIP Conference Proceedings}, 722(1):131--140, 2004.

\bibitem{WFMSGRID05}
Yu~J and Buyya R.
\newblock {A Taxonomy of Workflow Management Systems for Grid Computing}.
\newblock {\em Journal of Grid Computing}, 3(3):171--200, 2005.

\bibitem{WLCG06}
Bonacorsi D and Ferrari T.
\newblock {WLCG Service Challenges and Tiered architecture in the LHC era}.
\newblock In {\em IFAE 2006}, pages 365--368. Springer Milan, 2007.

\bibitem{ADC17}
{ATLAS} {Collaboration}.
\newblock {ATLAS Distributed Computing experience and performance during the
  LHC Run-2}.
\newblock {\em J. Phys. Conf. Ser.}, 898(5):052015, 2017.

\bibitem{PROD17}
{ATLAS} {Collaboration}.
\newblock {The ATLAS Production System Evolution: New Data Processing and
  Analysis Paradigm for the LHC Run2 and High-Luminosity}.
\newblock {\em J. Phys. Conf. Ser.}, 898(5):052016, 2017.

\bibitem{DDM17}
{ATLAS} {Collaboration}.
\newblock {Experiences with the new ATLAS Distributed Data Management System}.
\newblock {\em J. Phys. Conf. Ser.}, 898(6):062019, 2017.

\bibitem{TAPEPLACEMENT06}
Zhang X, He~D, Du~D Hc, et~al.
\newblock {Object Placement in Parallel Tape Storage Systems}.
\newblock In {\em 2006 International Conference on Parallel Processing
  ({ICPP}'06)}, pages 101--108, 2006.

\bibitem{STORAGECOST07}
Moore RL, D'Aoust J, McDonald RH, et~al.
\newblock {Disk and Tape Storage Cost Models}.
\newblock {\em Archiving Conference}, 2007(1):29--32, 2007.

\bibitem{YU_2011}
Yu~D and Lauret J.
\newblock {Tape Storage Optimization at BNL}.
\newblock {\em Journal of Physics: Conference Series}, 331(4):042045, 2011.

\bibitem{DATACAROUSEL20}
{ATLAS} {Collaboration}.
\newblock {ATLAS Data Carousel}.
\newblock Technical Report ATL-SOFT-PROC-2020-014, CERN, Geneva, 2020.

\bibitem{vrogcp}
Lim K.
\newblock {DMTN-125: Google Cloud Engagement Results}.
\newblock \url{https://dmtn-125.lsst.io/}.
\newblock Accessed: 2020-11-13.

\bibitem{dataocean}
{ATLAS} {Collaboration}.
\newblock {The Data Ocean project: An ATLAS and Google R\&D collaboration}.
\newblock {\em EPJ Web Conf.}, 214:04020, 2019.

\bibitem{lhcopn}
Martelli E and Stancu S.
\newblock {LHCOPN} and {LHCONE}: {Status} and {Future} {Evolution}.
\newblock {\em J.Phys.Conf.Ser.}, 664:052025, 2015.

\bibitem{gcp}
Google.
\newblock {Google Cloud Platform}.
\newblock \url{https://cloud.google.com/}.
\newblock Accessed: 2020-09-14.

\bibitem{prodsys}
{ATLAS} {Collaboration}.
\newblock {Multilevel Workflow System in the ATLAS Experiment}.
\newblock {\em J.Phys.Conf.Ser.}, 608:012015, 2015.

\bibitem{panda}
{ATLAS} {Collaboration}.
\newblock {Overview of ATLAS PanDA workload management}.
\newblock {\em J.Phys.Conf.Ser.}, 331:072024, 2011.

\bibitem{cms_caching}
Marco Meoni, Raffaele Perego, and Nicola Tonellotto.
\newblock Dataset popularity prediction for caching of cms big data.
\newblock {\em Journal of Grid Computing}, 16(2):211--228, Jun 2018.

\bibitem{atlas_caching}
M.~Titov, G.~Zaruba, et~al.
\newblock {A probabilistic analysis of data popularity in ATLAS data caching}.
\newblock {\em J. Phys. Conf. Ser.}, 396:032106, 2012.

\end{thebibliography}

\end{document}